# METALLIC PHASE OF HYDROGEN AT THE PRESSURE 500 GPA

N. Degtyarenko and E. Mazur

National Research Nuclear University "MEPHI", Kashirskoe sh.31, 115409 Moscow, Russia

eugen_mazur@mail.ru

**Abstract**. The first-principle method of mathematical modeling was used to calculate the structural, electronic, phonon, and other characteristics of the normal metallic phase of hydrogen at a pressure of 500 GPa. It has been shown that metal hydrogen having a lattice symmetry I41 / AMD cell is a stable phase under a high pressure hydrostatic compression. The resulting structure is a locally stable one with respect to the phonon spectrum.

**Introduction**

In the theoretical paper [1] in the study of the properties of metallic hydrogen very high critical superconducting temperature Tc ~ 200 - 400 K is predicted. Moreover, in accordance with the calculations of Brovman E.G. and Kagan Yu. [2, 3] atomic hydrogen may form metastable phase under reduced pressure up to the atmospheric pressure values. The reaching the superconducting state in a metal atomic hydrogen or in the hydrogen-containing compounds would be important for understanding the mechanism of high temperature superconductivity and may open the road towards room temperature superconductivity. However, the conversion of molecular hydrogen into a metal requires huge pressure about 400-500 GPa, which is higher than the pressure value achievable at present in the art of high pressure. Carrying out ab-initio calculations to predict the pressure region in which there is synthesis of metallic phase of hydrogen, the study of its stability and the study of the properties of the metal hydrogen phase at high pressures is an urgent task. In this respect, there is a large number of studies [4-10] at present. In this study, we discuss the properties of a number of hydrogen phases at a pressure of 500 GPa. The phase having properties of metallicity, the phonon spectrum of which contains no imaginary frequencies at a specified pressure is found.

**Method**

The calculation of the structural, electronic and phonon characteristics of molecules as well as of the periodic crystal structure of hydrogen at a pressure of 500 HPA is carried out by ab-initio. The approach in the framework of DFT basis of plane waves with the correlation functional GGA - PBE with pseudopotential that preserves the norm is used. All the calculations were performed in the spin-polarized approximation for adequate comparison of the calculated values of the energy of the different phases. The level of "circumcision" of the kinetic energy is increased up to 2721 eV in accordance with the high values of the system compression. The threshold of the self-energy convergence in the optimization of the system configuration was set to $5 \times 10^{-6}$ eV / atom. The method of the crystal super cell with periodic boundary conditions is used to simulate the properties of the phase in study. In the calculations of the phonon spectra DFPT method is used with an expanded number of unit cells due to the small concentration of electrons.

**Results**

The calculations of several crystalline phases of hydrogen are performed for the pressure P = 500GPa. The structures produced as a result of optimization of the geometry of the lattice with the principle of minimizing the total energy are shown in Fig. 1 (*a, b, c*). The resulting symmetry of these structures should be marked as I4 / MMM, P6 / MMM, I41 / AMD:

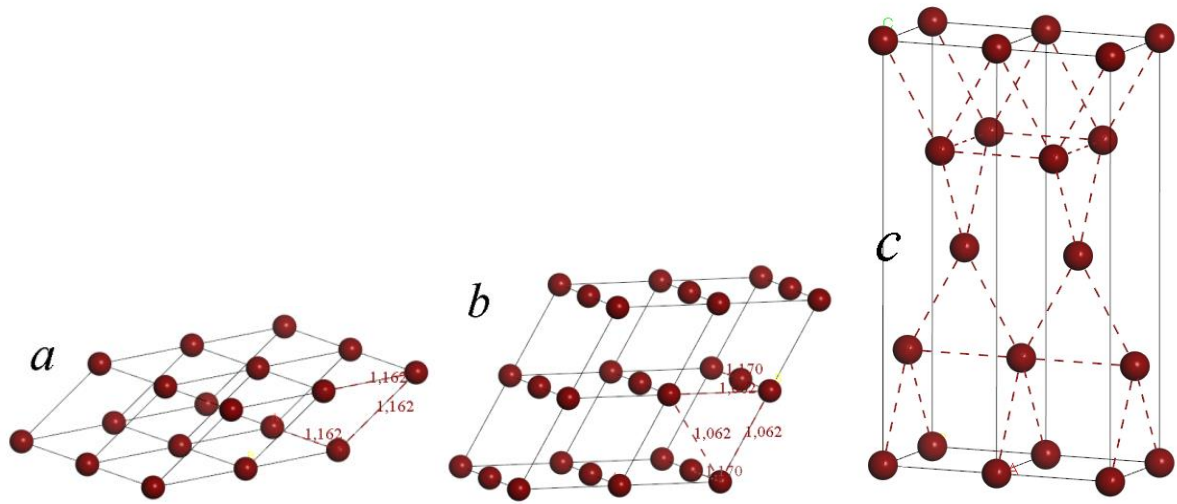

Figure 1 (a, b, c). The structures of the crystalline phases of metallic hydrogen with different symmetry under hydrostatic pressure P = 500GPa

The values of the enthalpy H of these phases are very close one to another as follows: -9.662; -9.740; -9.819 [eV / atom]. The specific value of the volume per atom also differs slightly, namely: 1.128; 1.143; 1,149 [$A^3$/ atom]. The loose phase has the lowest calculated value of the enthalpy with respect to other phases. Note that the second structure is similar to the poorly defined layered graphite-like lattice. The third structure was previously considered in [4]. Despite the small difference of the enthalpies $\Delta H \approx 0.1 \div 0.2$ eV / atom of these structures, the phonon dispersion relations of the structures vary greatly. The results of the calculations of the phonon spectra are shown in Fig.2. From Fig. 2, it follows that the first two structures are unstable, since there are imaginary frequency (negative values) in their dispersion dependences. The third structure is stable, and its maximum frequency reaches $h\nu_{max} \approx 340$ meV. Note that if we consider the hydrogen atoms as the quantum harmonic oscillators, and the barriers between the different phases as the difference between the enthalpy values $\Delta H \approx 200$ meV between these phases, then $\Delta H$ is shown to be less than $3/2 * h\nu_{max}$. Next we consider the properties of only the structure shown in Fig. 1. *c*. The primitive cell of this structure and of its reverse cell is shown in Figure 3. The cell contains two hydrogen atoms. The number of nearest neighbors equals three. The equilibrium distance for 500GPa pressure equals $r_0$ =0.986A. There are two Fermi surfaces in the reciprocal space marked with the different colors as is shown in Fig. 4.

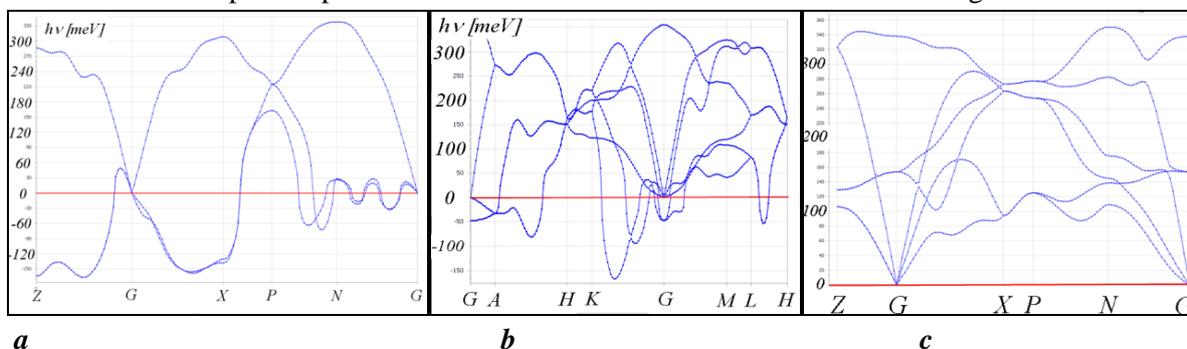

*a*            *b*            *c*

Fig.2. Phonon dispersion relations for the structures of atomic hydrogen phases presented in Fig. 1.

All of the phases have a metallic character. Fig. 5 shows the electron dispersion curves and the partial density of electronic states PDOS, which refers to the s-electrons. The Fermi level crosses two bands and is close to the minimum density of electronic states.

The electron density distribution in the real space is shown in Fig. 6.

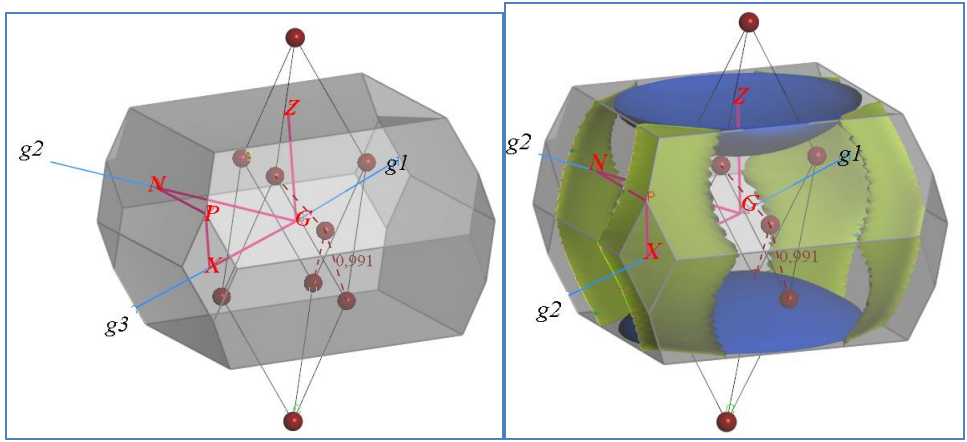

Figure 3. Primitive cell of a structure with symmetry I41/AMD metallic hydrogen at a pressure of P=500GP

Figure 4. The Fermi surfaces in reciprocal space for the structure with symmetry I41/AMD metallic hydrogen at a pressure of p=500GP

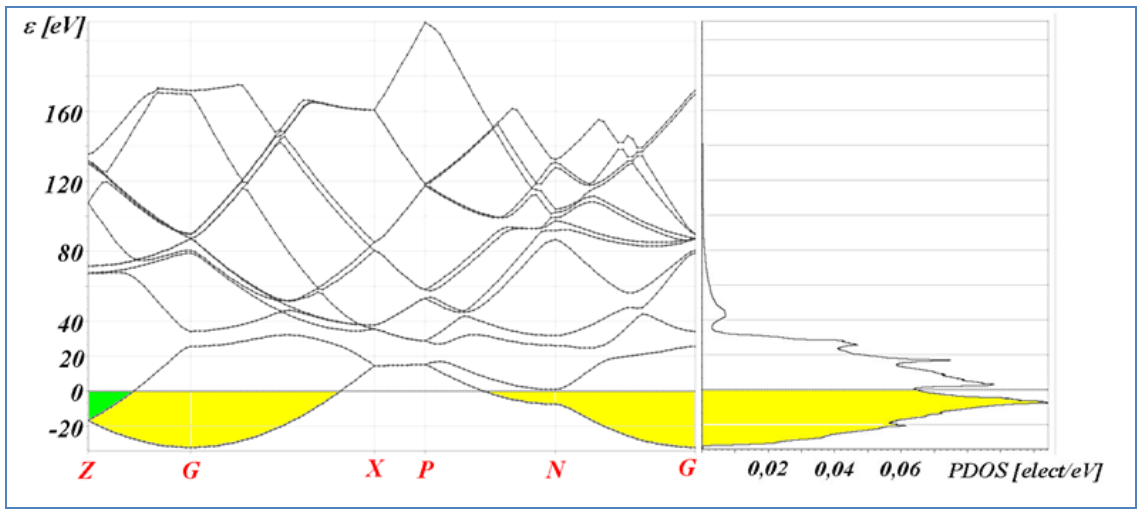

Figure 5. Dispersion curves of electrons and the partial number density of s–electron states PDOS for the structure with symmetry I41/AMD metallic hydrogen at a pressure of P=500GP

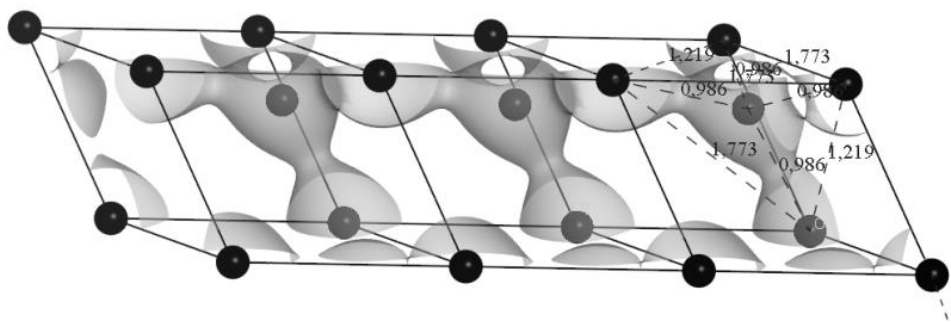

Figure 6. The isosurfaces of electronic density at the level of 1 electron per $A^3$ for the structure with symmetry I41/AMD metallic hydrogen at a pressure of P=500GP. The minimum value of the electron density in the space between the atoms is of the order of 0.5

## Conclusions

At a pressure of 500 GPa the structure with the symmetry I41 / AMD is shown to be the stable metallic phase of atomic hydrogen. The phonon spectrum of this phase does not contain imaginary frequencies and extends to a maximum frequency of about 340 meV. This structure has a lower enthalpy at the given pressure value P as compared with the other two calculated phases, but the difference between the enthalpy values is small and amounts the value of the order of 0.2 eV.